\documentclass[sigconf]{acmart}
\usepackage{pgfplots}
\pgfplotsset{compat=1.16}

\AtBeginDocument{%
  \providecommand\BibTeX{{%
    \normalfont B\kern-0.5em{\scshape i\kern-0.25em b}\kern-0.8em\TeX}}}

\copyrightyear{2022}
\acmYear{2022}
\setcopyright{rightsretained}

\acmConference[CHI '22]{CHI Conference on Human Factors in Computing
Systems}{April 29-May 5, 2022}{New Orleans, LA, USA}
\acmBooktitle{CHI Conference on Human Factors in Computing Systems
(CHI '22), April 29-May 5, 2022, New Orleans, LA, USA}
\acmDOI{10.1145/3491102.3517716}
\acmISBN{978-1-4503-9157-3/22/04}

\begin{document}

\title{A Systematic Review and Thematic Analysis of Community-Collaborative Approaches to Computing Research}

\author{Ned Cooper}
\orcid{0000-0003-1834-279X}
\affiliation{%
  \institution{School of Cybernetics\\
  Australian National University}
  \city{Canberra}
  \country{Australia}
}

\author{Tiffanie Horne}
\affiliation{%
  \institution{Google LLC (via Pro Unlimited)}
  \city{Seattle}
  \country{USA}}

\author{Gillian R. Hayes}
\orcid{0000-0003-0966-8739}
\affiliation{%
  \institution{Department of Informatics\\
  University of California, Irvine}
  \city{Irvine}
  \country{USA}}

\author{Courtney Heldreth}
\affiliation{%
 \institution{Google}
 \city{Seattle}
 \country{USA}}

\author{Michal Lahav}
\affiliation{%
  \institution{Google}
  \city{Seattle}
  \country{USA}}

\author{Jess Holbrook}
\affiliation{%
  \institution{Google}
  \city{Seattle}
  \country{USA}}
  
\author{Lauren Wilcox}
\orcid{0000-0001-6598-1733}
\affiliation{%
  \institution{Google Research}
  \city{Mountain View}
  \country{USA}
}

\renewcommand{\shortauthors}{}

\begin{abstract}
HCI researchers have been gradually shifting attention from individual users to communities when engaging in research, design, and system development. However, our field has yet to establish a cohesive, systematic understanding of the challenges, benefits, and commitments of community-collaborative approaches to research. We conducted a systematic review and thematic analysis of 47 computing research papers discussing participatory research with communities for the development of technological artifacts and systems, published over the last two decades. From this review, we identified seven themes associated with the evolution of a project: from establishing community partnerships to sustaining results. Our findings suggest that several tensions characterize these projects, many of which relate to the power and position of researchers, and the computing research environment, relative to community partners. We discuss the implications of our findings and offer methodological proposals to guide HCI, and computing research more broadly, towards practices that center communities.
\end{abstract}


\maketitle

\section{Introduction}
HCI researchers have long recognized the value of participatory methodologies in computing research, embracing approaches with roots in Action Research (AR) \cite{Hayes2011-be} and Participatory Design (PD) \cite{Muller1993-tm} for their democratic potential to open the design process to input from the intended `users' of technology. More recently, HCI has increasingly adopted complementary approaches to PD that can broadly be described as Community-Based Participatory Research (CBPR). Approaches based on CBPR aim to bring communities of expected `users' more collaboratively into the development of technological artifacts and systems, and to do so throughout the research process \cite{Israel2012-je, Wallerstein2006-he}. Yet, as we describe in more detail below, scholarship on the nature of community collaboration remains largely siloed from scholarship in HCI and computing research more broadly. Without sufficient grounding in the relevant body of work, there is a risk that computing-focused research may exploit and extract from communities without offering mutual benefits or attending to fundamental inequities \cite{Sloane2020-pu, Pierre2021-rs}.

As a multidisciplinary field, HCI is well placed to reflect on the nature of community collaboration in computing research to-date, and offer guidance for future projects. We thus conducted a reflexive thematic analysis of prior work identified through a systematic review, which enabled us to identify, review, and reflect on findings from projects that adopt AR, CBPR, and related approaches to computing research.  To be inclusive of the disciplines that contribute to these approaches without narrowing to any one, we refer to them under the umbrella of ``community-collaborative approaches'' or CCA, a term not in use in the existing research literature. We intentionally chose to develop a new term to encompass approaches that might co-exist together, enabling us to examine a variety of methods and approaches collectively that hold in common these features: participation (as research team members) of people from the community of interest, not just for formative or evaluative feedback, but throughout research and development. We focus this literature review and analysis on computing research, broadly, and draw on the analysis to reflect on disciplinary norms of computing research, distilling from these reflections specific recommendations for  researchers. 

Contributing disciplines to CCA, described further in the Background, each hold particular epistemic positions, values, and ways of conceptualizing and reporting on research. This paper pursues answers to the overarching question: What do \textit{researchers} publishing computing literature talk about when talking about community-collaborative research? Specifically, three key research questions guided our review:
\begin{enumerate}
\item How do computing researchers describe their efforts to involve community stakeholders as co-researchers throughout collaborative research projects?  For which types of projects and with which communities do we see collaborative research published in computing literature?
\item What elements of community collaboration are reported in computing research publications, and what are the cultural and methodological implications of foregrounding those elements?
\item How does the content in these publications align with or diverge from the tenets of Action Research, Community-Based Participatory Research, and similar `community-collaborative' approaches that center community participation? What critiques are suggested by divergence, and what methodological innovations could move forward research in both HCI and computing more broadly, to attend to these critiques?
\end{enumerate}

This research contributes the following to the CHI community:
\begin{enumerate}
\item A meta-review of how computing researchers have adopted community-collaborative approaches, to enable HCI researchers to learn from a spectrum of prior projects.
\item A reflexive thematic analysis of research strategies, challenges, and project outcomes, as reported in prior work, through which we identify seven themes that reflect dimensions of community-collaborative research, as well as tensions that emerge across stages of a project.
\item A reflection on our findings and methodological proposals to enhance the commitment of our field to community-collaborative approaches, distilling opportunities for new areas of research in HCI. 
\end{enumerate}

This paper provides a starting point for researchers who plan to engage with community stakeholders to develop technological artifacts and systems. In the following sections, we provide background on the emergence of CCA in computing research to date and critical perspectives on its use in technology design. We then outline our protocol for selecting papers and our process for reflexive thematic analysis. Next, we present and describe the material we reviewed, including the types of communities involved and the types of projects conducted. We report findings on seven themes interpreted from the paper corpus associated with the evolution of a project. Our findings suggest that several tensions characterize the projects, requiring close attention and specific commitments by researchers, both during the course of the project and beyond. Many of the tensions require researchers to be aware of and address their power and position, and that of the computing research environment, relative to community partners. Finally, we offer provocations related to structural changes needed in our research communities to support engagement with broader communities, methodological norms that can support CCA, and the necessity for---and challenges to---implementation of reflexive practice in community-collaborative research. In doing so, we provide a foundation for research teams to facilitate the meaningful and mutually-beneficial participation of communities in HCI and computing research.

\section{Background}
\subsection{HCI and Action Research}
The past twenty years of HCI research have marked an increase in efforts connecting significant societal concerns with technology design and research. Hayes charts the progress of this increase, describing the upward trend in CHI papers focused on ``civically-engaged'' research since 1990, with particular acceleration post-2005 \cite{Hayes2011-be}. In many ways, such work aligns with the participatory stance of AR, a class of ontological, epistemological, and methodological approaches for conducting collaborative research with community partners in ways that aim to democratize the research process.

AR originally developed in the late 1930s, as an approach to apply ideas from social psychology to practical real-world problems. For example, original research studies by graduate students of Kurt Lewin, who immigrated to the US during WWII and who is credited with coining the term, found greater gains in productivity and civic adherence to laws through ``democratic participation rather than autocratic coercion'' \cite{Adelman1993-ca}. Though it has many incarnations and is not limited to sociotechnical domains, AR offers a compelling, systematic approach for sociotechnical researchers, as it emphasizes collaboration with community stakeholders to co-construct knowledge and apply it to real-world social, structural, material, and environmental phenomena \cite{Lewin1946-jy,McTaggart1991-hf, Hayes2011-be, Torre2012-jg, Avison1999-de}.

In the 1960s, AR began to influence sociotechnical systems thinking in European organizations and organizational research \cite{thorsrud1977democracy}. Later in the 1970s, advocates of the workplace democracy movement in Scandinavia drew on AR-inspired approaches to respond collectively to the introduction of computer systems in the workplace \cite{Kensing1998-tk}. Trade unions and designers advocated for direct worker participation in the design and decision-making about such systems, to limit the deskilling and devaluing of human work \cite{DiSalvo2013-ee, jarvinen1981user}. PD emerged from these early experiments in Scandinavian workplaces, into a method for combining the practical and experiential knowledge of `users' with the conceptual knowledge of researchers and designers \cite{Sanders2008-oo,Muller2012-tq,Bjorgvinsson2010-gk}.

As the shift to PD emerged in Europe, it became gradually more established as a distinct, democratic approach to involving, early on, the people who would be impacted by the design of a (sociotechnical) system. Though it was aligned with tenets of AR, PD soon came to be recognizable as its own disciplinary movement. As the ACM CHI conference began in the early 1980s, scattered references to PD could be found in the broader computing literature, and by the early 1990s, PD had been introduced in CHI panels \cite{Johnson1990-jj} and papers\cite{blomberg_reflections_1990}, becoming increasingly applied to original work in HCI throughout the 1990s, with gradually more adoption over time. 

Though related in many ways to AR, PD offered its own lessons, case studies, and methodological insights which were, in turn, shaped by the particular ways in which HCI adopted PD. Thus, the original influence of AR on PD shifted somewhat as HCI adopted and contributed to its evolution as an applied design research approach. Many PD projects conceptualized participation as the involvement of individuals---often within formal organizations---with varying forms of creative input and research involvement \cite{Sabiescu2014-ka}. In fact, many PD projects still engage individual-level feedback to inform design decisions, including in design workshops, which are often removed from local community sites and rely largely on researcher identification of individual participants \cite{Rosner2016-dd,Sloane2020-pu,Donia2021-bb}.

Over time, PD also moved from a workplace-specific approach to one used in a variety of other contexts and named with other labels. For example, Xie et al. explored working with children and older adults as part of co-design teams \cite{xie2012connecting} while Fredericks et al. developed ``middle out'' design approaches for urban development \cite{fredericks2016middle}. More recently, researchers have explored collaborative and PD projects led by historically marginalized groups, such as Indigenous people \cite{Peters2018-js} and Black Americans \cite{harrington2021eliciting}. Tracing this evolution, one can see not only a change in stance towards participation but also an evolution in methods, including moves out of laboratories and design studios into community spaces and even remote design, partially accelerated by the COVID-19 pandemic.

\subsection{The Emergence of Community-Based Approaches in HCI}
CBPR, ``community-based'' research, ``community-led'' research, and a variety of other similar names describe a set of approaches that have emerged in social and health sciences to address societal needs through the participation of community members as equal partners in research \cite {murphy1984community, baum2006participatory, feldman1974bureaucratic}. These community-collaborative research approaches were gaining momentum in the social sciences, humanities, and public health in parallel to the rise of PD in HCI in the 1990s–2000s, specifically for engaging marginalized communities in research. As noted in the Introduction, we use the term CCA (community-collaborative approaches) for the umbrella of work in HCI, and computing research more broadly, that includes AR, CBPR, and related efforts, to involve collaboration with community stakeholders as co-researchers throughout the research process, while investigating complex problems of concern to the community (\textit{e.g.}, social and health-related), and developing novel technologies or sociotechnical solutions. These projects typically have a goal of bringing about change for particular communities \cite{Boyd2014-hq,Horowitz2009-zl}. 

As detailed in our full review below, by 2010, a handful of CHI papers reported using CCA to guide participatory research projects. Over the past decade, more PD projects published in HCI and computing venues have involved participation of marginalized communities in aspects of research. At the same time, others have begun to critique the use of PD in HCI projects with marginalized communities \cite{Erete2018-qr}. Critiques have interrogated the social and cultural positioning of the typical PD workshop \cite{Harrington2019-yc, Irani2018-yj}, and the focus in PD on data collection that primarily serves researchers for the purpose of developing a design intervention \cite{Racadio2014-gi}, rather than enabling the community to enact social change as part of the research process \cite{akom2016youth, Harrington2019-ze}. Notably, although these critiques have recently entered the discourse in HCI communities, such concerns have existed for decades in AR, CBPR, and PD scholarly discussions (\textit{e.g.}, \cite{grant2008negotiating, frisby1997reflections, asaro2000transforming, ertner2010five, depaula2004lost, bodker.kyng_2018}).

The canon of acceptable methodological approaches is continually evolving in HCI, one of the hallmarks of this interdisciplinary, innovative, and iterative field. The inclusion of notions of design justice \cite{Costanza-Chock2020-iz} and design for social justice \cite{dombrowski2016social} is one such evolution. Design justice emerged from a growing community of practitioners who partner with social movements and community-based organizations to ``ensure a more equitable distribution of design’s benefits and burdens; meaningful participation in design decisions; and recognition of community-based, Indigenous, and diasporic design traditions, knowledge, and practices'' \cite{Costanza-Chock2020-iz}. A design justice framework illuminates how design can reproduce or challenge structural inequality, with the goal of guiding design practitioners to avoid reproducing existing inequalities, and enabling communities to use design practice and process to facilitate liberation from oppression \cite{Costanza-Chock2020-iz}.

The design justice framework is a natural analog to the CCA we discuss in this paper, moving from educational, social, and behavioral disciplines into computing research. CCA and design justice share commitments to foregrounding engagement with marginalized communities and co-constructing knowledge through participatory activities. In CBPR, for example, researchers and community stakeholders work together in all stages of a project, to investigate problems identified by, and valuable to, the community, and then combine knowledge and action for social change \cite{Israel1998-yg,Minkler2011-hm}. Community-collaborative projects strive to cultivate embedded community partnerships, foreground the community's collective voice, and engage with their epistemologies alongside those of the researchers. In this way, CCA research must be conducted \textit{for}, \textit{with} and \textit{by} communities rather than \textit{on} communities \cite{Fox2017-ss,Koster2012-jj,DiSalvo2013-ee}.

CCA can increase the relevance of products and services to specific \textit{groups} of people as well as unearth potential negative impacts before they become issues. These approaches focus on community influence on the technology development process. By flipping this influence from an intervention impacting the community to a community impacting the intervention, the creativity and innovation of that process grows. CCA recognize that the lived experiences of community members bring unique strengths and resources to bear on technological innovation as well as policies and practices surrounding novel technologies. Effective deployment of these resources can mitigate the potential for current and emerging technologies to have disparate or potentially harmful impacts on communities who already experience marginalization.\footnote{Marginalized communities are defined in this paper as communities experiencing exclusion from mainstream social, political, economic, and cultural life because of unequal power relationships \cite{Baah2019-je,Vasas2005-ux}. The term `marginalized' is used throughout this paper instead of `underserved', `underrepresented', or `vulnerable'. Our paper reflects the view that the term `marginalized' better reflects ``a failing of society, rather than a failing of any individual person'' \cite{Liang2021-qe}. Nonetheless, we acknowledge it can be problematic to ascribe any term to a community from outside the community \cite{Williams2020-fe}. As such, whenever possible in this paper we reflect names used by members of a community to describe their community.}

\subsection{Debates About Participation}

The development of PD since its early focus on ``democracy at work'' has been accompanied by debates as to whether PD projects continue to shift power from those designing technology to those who will use technology \cite{Bjorgvinsson2010-gk}. As Muller and Kuhn \cite{Muller1993-tm} ask, who participates with whom in what? Do researchers and designers participate in the world of `users', or vice versa? If the latter, what forms of `user' knowledge are accepted in the world of researchers and designers? Additionally, PD scholars have asked who benefits from the participation of `users' in HCI research---those who use technology or those who seek to improve a product or service for use by others? \cite{Vines2013-yf}.

More broadly, as HCI has grown from a field focused on ergonomics and human factors interests of workers to an interdisciplinary space concerned with a wide array of technology uses and the social phenomena surrounding use, the notion of who participates and how they participate has a wide range of consequences. For example, in the subfield of HCI interested in assistive technology, authors have argued for much greater participation of people with disabilities in research about them \cite{spiel2020nothing} while others have pushed back against ``identity politics'' \cite{kirkham2021disability}. As a research community we are beginning to problematize the very notion of identity and inclusion in our discussions of participation and community. We are still examining what it means to operationalize positions of inclusion that require, for example, someone with a disability, or part of any community of interest, to participate at a level visible to the research community (\textit{e.g.}, as an author or named collaborator) on every paper published in a space. An extreme version of such a position could require at least one author on every paper to `out' themselves as being a part of the community, perhaps a stigmatized community, even if they do not want to. Such a position could lead to the unintended consequence of tokenizing people from those communities and their inclusion in the publication of papers.\footnote{We note that some mechanisms for publishing without identification, such as the ACM pseudonymous publishing policy, do exist and could be expanded.} Thus, while we advocate for the full participation of community members to the degree that is useful, interesting, and productive for them, `participation' is not an unqualified good. 

There are no easy answers to debates about what it means to participate in HCI research projects nor to collaborate with HCI and other computing scholars as a community member from outside the world of scholarship. Transparency and communication can better enable community members to make informed choices, but these choices will always be laden with other concerns, such as the resources that might be tied to participation or the moral imperative to be a part of research that might impact their community. For HCI and related computing researchers, for and about whom this paper is primarily written, we sought to understand what the additional responsibilities of CCA might be relative to other research approaches, including PD. We seek in this paper to interrogate the nature of community-collaborative computing projects, reflecting on the past two decades of practice as represented through the discourse available in scholarly research publications. This analysis provides insight into how projects have unfolded that allows us to learn from the experiences of other researchers and to identify areas calling for the attention of future HCI researchers to fulfill the aspirations of CCA.

\section{Method}
Our method combined a systematic review for data collection and a reflexive thematic analysis for data analysis. The primary contribution of this paper is the reflexive thematic analysis, though we outline our approach to both data collection and analysis in detail below. We conclude the section with a reflection on the positionality of the research, to add important context to our analyses and findings, and a discussion of the limitations of our approach.

\subsection{Data Collection}

Our approach to data collection followed prior systematic reviews in HCI literature, including reporting in accordance with the Preferred Reporting Items for Systematic Reviews and Meta-Analyses (PRISMA)  \cite{Harris2019-ev,Caraban2019-ys,Hansson2021-in,Hollander2021-lh,Pater2021-ac, Jelen2020}. Data collection comprised four steps: database searches, abstract screening, full-text screening and eligibility checks, and corpus data extraction. Below, we report our methodology for each of these steps.

\subsubsection{Information Sources and Search Strategy}

Our search consisted of five databases: the ACM Guide to Computing Literature (`ACM'), IEEE Xplore digital library (`IEEE'), ScienceDirect, Taylor \& Francis Online and Wiley Online Library. We included the latter three databases to expand our searches beyond computing and engineering publications, to capture any projects based on CCA that focus on developing a technological artifact or system. We iteratively developed a boolean search string with AND/OR terms across variants and abbreviations of various CCA. The terms in the search string are intended to identify computing research projects involving the participation of communities rather than discrete individuals. We also focus on research that is community-based (or ``led'' or ``driven''), rather than research about a community or placed in a community. However, we note the distinction between PD, AR, and CBPR is not clear cut---all may have aspects that are community-based. Thus, we included papers in our review that refer to ``participatory design'', ``participatory research'' or ``participatory action research'' where researchers work with a community as a unit vs. individuals.

We did not include gray literature in our review. As such, work produced by communities themselves discussing projects based on CCA but published outside peer-reviewed journals or conference proceedings is excluded from the review. Production of such work is a cornerstone of the commitments community-based researchers and communities make to one another and would be important for future analysis. However, those works were excluded here because the intent of this article is to focus on how CCA is represented in scholarly venues, and specifically, to examine the discourse allowed, enabled, and supported by the scholarly publication norms of computing research. As noted in our positionality statement further below, we as authors have participated in a variety of research dissemination efforts outside this corpus and expect other HCI and computing researchers have as well. That these efforts are not recognized by organizations as having scholarly import is a notable concern (addressed in the Discussion) that also explains their omission, however limiting, in our review.

For ACM and IEEE, we searched our string throughout full text or metadata; for ScienceDirect, Taylor \& Francis Online, and Wiley Online Library, we searched the string in the titles, abstracts or keywords of texts AND searched for an additional string in full text or metadata (see Tables 1 and 2 for search terms). Our searches included peer-reviewed, English-language articles published between January 1, 2000 to July 31, 2021. We note our review has a limited view of studies published in non-English venues. The searches yielded 1397 total texts.

\begin{table*}
\begin{tabular}{|l|} 
 \hline
 \textbf{Full text and metadata:}\\ 
 \hline (``community'' AND (``based'' OR ``led'' OR ``driven'') AND (``research'' OR ``design'' OR ``PD'')) OR \\
 (``community'' AND (((``participatory'' AND (``research'' OR ``design'')) OR ``participatory action research'')) OR\\
 ``CBPR''\\
 \hline
\end{tabular}
\caption{ACM and IEEE search terms}
\label{table:1}
\end{table*}

\begin{table*}
\begin{tabular}{|l|} 
 \hline
 \textbf{Title, abstract, and keywords:}\\ 
 \hline (``community'' AND (``based'' OR ``led'' OR ``driven'') AND (``research'' OR ``design'' OR ``PD'')) OR \\
 (``community'' AND (((``participatory'' AND (``research'' OR ``design'')) OR ``participatory action research'')) OR\\
 ``CBPR''\\
 \hline
 \textbf{Full text and metadata:}\\ 
 \hline
 ``computing'' OR ``technology'' OR ``information system'' \\ 
 \hline
\end{tabular}
\caption{ScienceDirect, Taylor \& Francis Online, and Wiley Online Library search terms}
\label{table:2}
\end{table*}

\subsubsection{Screening and Eligibility}

To prepare the corpus for screening, we removed duplicates and any remaining texts not peer-reviewed. We included articles that describe or discuss projects drawing on CCA for developing a technological artifact or system. We used this definition to include cases referring to one or more stages in the process of developing a specific artifact, system, or element of a system---from designing, to building, to testing, to deploying---rather than papers offering general guidelines or commentary on the technology development process. We defined a technological artifact as any digital or material object developed by humans to achieve some practical end, and a technological system as a set of interconnected elements (artifacts) developed by humans to fulfil some function.

We excluded articles that:
\begin{enumerate}
    \item Did not discuss the development of a technological artifact or system.
    \item Did not include community members participating in the research project (i.e. research was only about a community or placed in a community)
    \item Used a technological artifact or system to facilitate CCA, but did not intend to develop a technological artifact or system from the research.
    \item Used CCA to evaluate the impact of an existing technological artifact or system on a community, but not to develop that artifact or system together with the community.
\end{enumerate}
\begin{figure}[ht]
  \centering
  \includegraphics[width=\linewidth]{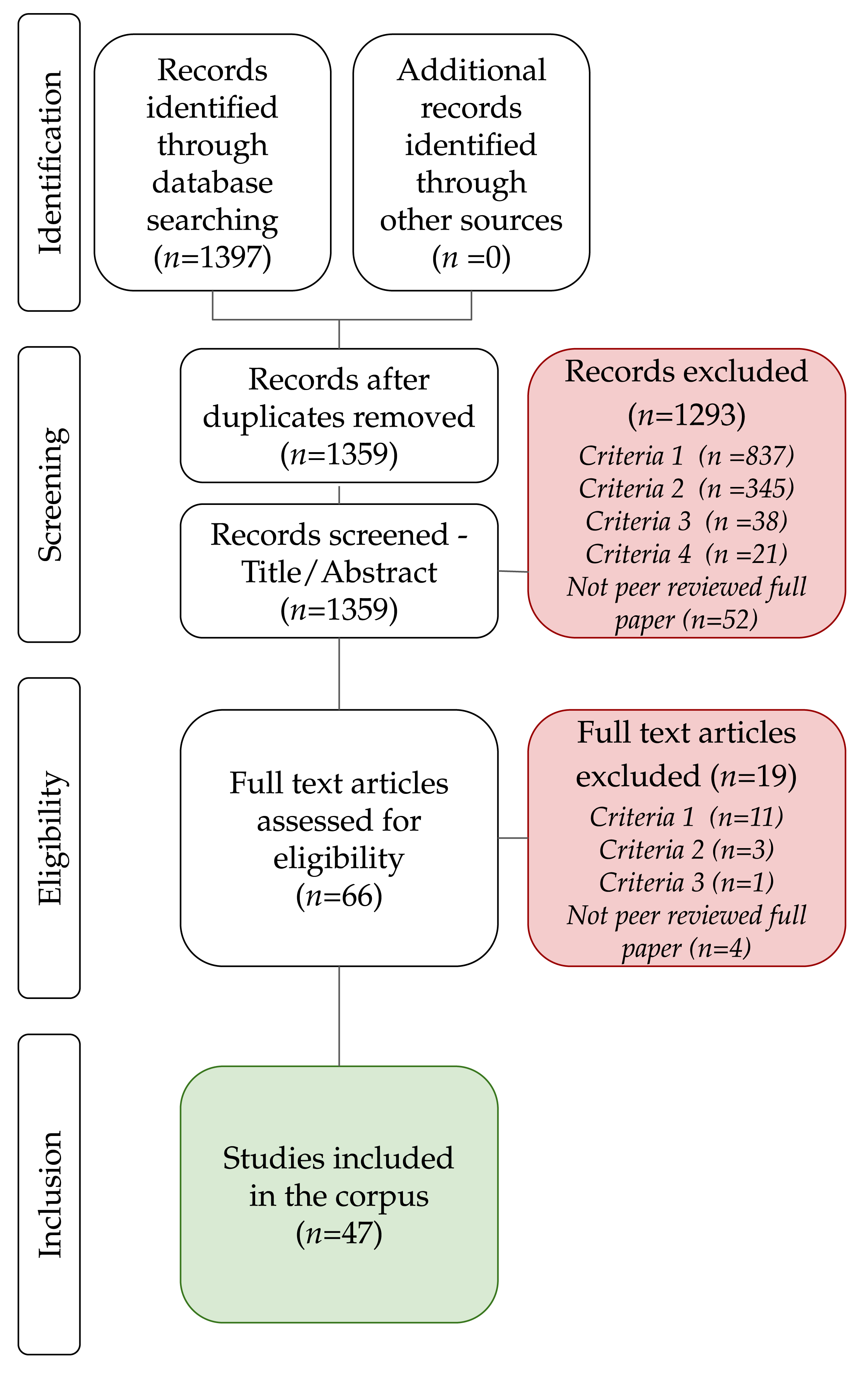}
  \caption{Adapted PRISMA flowchart outlining the process of selecting papers for inclusion in the corpus.}
  \Description{Flowchart outlining the process of selecting papers for review.}
\end{figure}
Two authors reviewed the titles and abstracts of a random sample of 50 articles and assigned `include', `exclude' and `unclear' labels, based on the eligibility criteria outlined above. To quantify agreement between the authors, we used Cohen's kappa coefficient \cite{Landis1977-bh}. The researchers achieved a consistent article selection ($\kappa\geq0.8$) and resolved disagreements for two articles in the sample through discussion. One author then reviewed the titles and abstracts of the remaining articles, using categorizations from the sample screening as a guide. After applying the criteria to all 1397 texts, 66 papers remained for full-text review (Figure 1).

Two authors independently read the full-text of each article and classified the articles as `include', `exclude' or `unclear'. The authors discussed and ultimately excluded 19 articles based on the eligibility criteria, leaving 47 articles for analysis (Figure 1).

\subsubsection{Data Extraction}

Next, we extracted quantitative and qualitative data from the 47 papers to conduct multiple analyses of the corpus. For each paper, we noted: (1) the distribution of papers per year, publication venue and geographical region; (2) the community participating in the research; (3) actors involved in the participatory practice (4); the stated design or research approach of the project; (5) the type of technological artifact or system; (6) the outcomes of the project; and (7) the time frame of the project. \footnote{Supplementary materials present this data}

\subsection{Data Analysis}

Our approach to data analysis drew on Braun and Clarke's guidelines for reflexive thematic analysis \cite{Braun2006-pq,Braun2020-mg}. Reflexive thematic analysis is a post-positivist approach to data analysis, recognising the influence of researchers in interpreting data and encouraging researchers to reflect on that influence as they develop and refine codes \cite{Braun2020-mg}. Such reflection helps research teams to recognize biases they bring to the analysis, but also to engage with the data holistically as a team of people with lived experiences that may be brought to bear on the analysis. Codes should evolve as the analysis progresses and researchers become more acquainted with the data (rather than pre-defining themes, then identifying the presence of themes in the data and testing for inter-rater reliability and related metrics) \cite{Braun2020-mg, byrne_worked_2021}. The approach is also collaborative, encouraging discussion among researchers undertaking the work to achieve rich interpretations of the data and to enable productive disagreements, rather than attempting to achieve consensus of meaning \cite{byrne_worked_2021}. Below we describe the flexible and organic process through which developed codes and finally generated themes from the data for this paper.

After initially engaging with the corpus during data collection, two authors each independently reviewed the full text of included papers. The authors used an inductive approach to openly develop codes based on the semantic content of paper in the corpus (in free-form, using spreadsheets rather than coding software). During code development, we engaged in deep and prolonged data immersion, discussion, comparison of interpretations of the meaning of the data, and refection. We met in multiple rounds to collaboratively revise and refine codes, including sharing disagreements in synchronous discussion. Finally, the two authors met in multiple rounds to reflect on data and relationships among codes to generate, revise and reflect on themes. We also discussed and recorded our categorizations of each article against the final set of themes in a summary spreadsheet.

The two authors then engaged in further discussions of explanatory memos with other authors, followed by drafting of content for this paper. We share the themes of CCA projects, based on the included articles, along with findings of our descriptive analysis in our Findings sections below.

\subsection{Positionality of the Research}

One goal of our review was to examine how computing research literature has conceptualized community participation to date. Importantly, as a review of papers published in scholarly venues, any critique thereby is not leveraged at the projects nor researchers themselves, but rather the entire system into which we collectively develop and disseminate scholarly contributions. The computing research environment, in which we are all complicit, describes and applauds CCA in particular terms, but individual researchers and their community partners may well describe projects differently outside of the scholarly discourse of publications. As such, some of the remedies we describe in the Discussion, as well as the limited scope of our searches noted earlier, relate to the need for a broader ability to discuss issues of `community' and `participation' in scholarly venues that are nontraditional. We note that we have ourselves published and presented on CCA in non-scholarly venues (\textit{e.g.}, conferences for practitioners, teacher education seminars, parent workshops, op-eds, and blogs). However, these labors go largely unrecognized at present by the scholarly research community and, as such, are not a part of the standard discourses at work in our research communities. We feel a kind of dissonance representing sometimes conflicting roles---as scholars who also serve as activists, allies, and practitioners---and the weight of responsibility that comes with serving as a translational bridge among varying groups.

Further, we note that the body of work we studied, by following traditional computing research paradigms typifying publication in computing research venues, conceptualizes CCA from a predominantly Western perspective. (Of the 47 publications in the corpus, 24 described projects situated in North America, six in Europe.) As such, we expect Western epistemology has shaped the representation of knowledge in the corpus. We aim to offer in this paper a better understanding of the positionality of such knowledge: how it conceives of roles of community members and computing researchers, how decision-making and collaborative ownership is defined and practiced, and what values and outcomes drive participatory projects. This synthesis should serve as a call for intentional pursuits of more diversified forms of knowledge and approaches to community participation in computing research as well as their broad distribution in scholarly venues and beyond.

The authors also acknowledge our own positionality inherits from these viewpoints—as computing researchers trained and working in predominantly Western institutions, we recognize the need for complementary scholarship to further the concepts and geographies studied in this paper. For example, many papers might use terms different from `community' or `participatory', but still make community-collaborative contributions to computing research. Our position also influences our subjectivity in inductive thematic analysis—we have not participated as \textit{community members} in participatory research projects, but instead as \textit{computing researchers}. Thus our perspective is imbued with relative societal privileges that many participants in projects based on CCA do not experience.

\subsection{Limitations}

We used systematic review and thematic analysis to integrate results from different but related qualitative studies published in academic venues. For qualitative studies, systematic reviews have an interpretative rather than aggregating intent, in contrast to systematic reviews of quantitative studies, and particularly when combined with reflexive thematic analysis. We recognise, however, that the breadth of data identified by the systematic review and the need to synthesize across studies limited the depth of our analysis and findings with respect to any one study or paper. A narrative review focusing on one or a few papers may be more useful for identifying and describing best practice in community-collaborative research projects. However, we suggest our synthesis approach provided the opportunity to uncover new insight into themes and trends \textit{across} community-collaborative research projects.

In addition, our findings include descriptive statistics to accompany each theme in the paper corpus. These statistics capture elements that papers \textit{reported}. We acknowledge a project might include more elements than those reported in the published literature we surveyed. For example, we provide descriptive statistics on the reporting of outcomes for communities and researchers, or lack thereof, though we acknowledge a project may have resulted in outcomes the authors do not report on.

\section{Findings: Descriptive Statistics}

In this section, we provide descriptive statistics of the included articles. In the next section we present seven themes generated from our inductive analysis of the articles, integrating additional descriptive data relevant to each theme. In the Discussion we critically reflect on the findings.

This section uses the terms `community members' to refer to all people who identify as part of a community, `participants' to refer to community members who participate in a project, and `researchers' to refer to HCI and computing researchers who work with a community to undertake a project. We acknowledge community members can take on many roles for a community-collaborative research project, but we distill these terms to differentiate between stakeholders referred to in the corpus.

Below, we discuss how articles varied in publication year, venue, geographical site of the research, community participants, types of technology resulting from the project, and time frame.\\

\textit{Publication Year}: Papers in the corpus were published across 14 of the 22-year span we surveyed. Almost two-thirds (\textit{n}=29, 62\%) were published in the last five years of this current study (2017--2021). Figure 2 provides the yearly distribution of papers in the corpus.\\ 

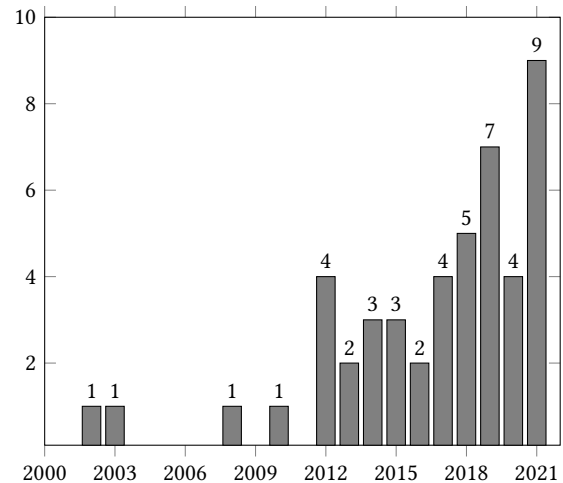
\begin{figure}[ht]
  \centering
  \begin{tikzpicture}
  \begin{axis} [ybar,
  xmin=2000,
  xmax=2022,
  ymax=10,
  nodes near coords,
  bar width=7pt,
  xtick={2000, 2003, 2006, 2009, 2012, 2015, 2018, 2021},
  xticklabels={2000, 2003, 2006, 2009, 2012, 2015, 2018, 2021}]
  \addplot [color=black,fill=gray] coordinates {
    (2002,1) 
    (2003,1) 
    (2008,1) 
    (2010,1)
    (2012,4) 
    (2013,2) 
    (2014,3)
    (2015,3) 
    (2016,2) 
    (2017,4) 
    (2018,5)
    (2019,7) 
    (2020,4) 
    (2021,9)};
    \end{axis}
    \end{tikzpicture}
  \caption{Number of articles (y axis) included in the corpus, by publication year (x axis).}
  \Description{Bar chart presenting distribution of publications across a 22-year period}
\end{figure}

\textit{Publication Venue}: Articles in the corpus appeared most commonly in the ACM CHI Conference (\textit{n}=8), ACM Conference on Computer Supported Cooperative Work and Social Computing (\textit{n}=6), ACM Designing Interactive Systems Conference (\textit{n}=5), Participatory Design Conference (\textit{n}=4), Australian Conference on Human-Computer Interaction (2), Conference on Information and Communication Technologies and Development (\textit{n}=2), African Human-Computer Interaction Conference (\textit{n}=2), and the ACM Transactions on Computer-Human Interaction (\textit{n}=2).\\

\textit{Geographical Location}: About half of the articles report on projects situated in the U.S. and Canada (\textit{n}=24, 51\%). Some articles referred to projects situated in the Global South (\textit{n}=13, 28\%), Europe (\textit{n}=6, 13\%), and Australia (\textit{n}=3, 6\%). Those in the Global South included projects situated in Africa and Asia. One article did not specify geographic location.\\

\textit{Community}: Table 1 presents the number of included articles that directly involved each community included in the corpus. Some articles relate to more than one community category (\textit{e.g.}, some relate to a `Rural' and `First Nations' community). A range of terms were used to refer to some community categories, and examples of these terms are included in the table for reference. The authors of this paper have attempted to reflect terms as reported in the papers.\\

\begin{table*}
  \caption{Communities, in descending order of inclusion in papers in the corpus. For rows without examples, the term in the \textit{Community} column is the term used in papers to describe the community.}
  \label{tab:freq}
  \begin{tabular}{lll}
    \toprule
    Community&Examples of How Described in Papers &Related Papers\\
    \midrule
    First Nations & `Indigenous', `Native American'& \cite{Duarte2021-cz,Kapuire2015-ou,Keskinen2021-ce,Winschiers-Theophilus2013-xj,Vigil-Hayes2021-rs,Sabiescu2014-ka,Peters2018-js,Darroch2017-tm,Blake2021-hg,Jensen2012-ww}\\
    People of Color in the U.S. & `Black', `African American' & \cite{Harrington2019-yc,Dickinson2021-mv,OLeary2020-qw,Harrington2019-ze,Stowell2020-yf,Le_Dantec2015-zk,Ginossar2010-uj,Mehra2002-jr,Parker2012-sf}\\
    People with a Disability & `Visually impaired', `People with dementia' & \cite{Neto2021-jf,Baldwin2019-eu,McNaney2018-mv,Grond2019-vw,Moharana2019-ra,Kossyvaki2018-jg,Khan2021-ug,Simm2014-rw,Zhu2019-ak}\\
    Rural community & & \cite{Sabiescu2014-ka,Ssozi-Mugarura2016-on,Duarte2021-cz,Kapuire2015-ou,Keskinen2021-ce,Winschiers-Theophilus2013-xj,A_Zewge2015-hr,Jensen2012-ww}\\
    Low-income community & `Poor community' & \cite{Ginossar2010-uj,Harrington2019-ze,Harrington2019-yc,Stillman2012-dd,Fox2014-ko}\\
    Residential community & `Local community' & \cite{Hsu2017-sh,Aoki2017-nl,DiSalvo2012-rs,Grabill2003-ha,Shilton2008-jk}\\
    Urban community & & \cite{Fox2014-ko,Harrington2019-yc,Stillman2012-dd,Darroch2017-tm,Parker2012-sf}\\
    Older Adults & & \cite{Tandukar2021-db,McNaney2018-mv,Harrington2019-ze,Harrington2019-yc}\\
    African community & `Ethiopian community' & \cite{Stillman2013-hg,Wojciechowska2020-jv,A_Zewge2015-hr}\\
    Health and Outreach Workers & `Community Health Workers' & \cite{Dickinson2021-mv,Thinyane2018-gt,Molapo2016-nc}\\
    Refugees and Forced Migrants & & \cite{Xu2017-cg,Duarte2018-iq}\\
    Transgender and Non-Binary & & \cite{Ahmed2021-ew,Haimson2020-ip}\\
    Women & & \cite{Darroch2017-tm,Mehra2002-jr}\\
  \bottomrule
\end{tabular}
\end{table*}

\textit{Stated Research Approach}: The articles referred to 13 different approaches. For simplicity, these have been grouped into two categories below:
\begin{itemize}
    \item Community-based research approaches:\\
    ``Community-based research'', ``Community-based design'', ``Community-based design research'', ``Community-based co-design'', ``Community-based collaborative design'',\\
    ``Community-based participatory research'', and ``Asset Based Community Development''.\\
    \item Participatory research approaches:\\
    ``Participatory design'', ``Co-design'', ``Collaborative design'', ``Co-inquiry'', ``Participatory Action Research'', ``Action Research''.
\end{itemize}

The majority of articles referred to a participatory research approach while describing \textit{elements} of the project that were community-based (\textit{n}=27, 57\%). Others referred to a community-based research approach, signalling the involvement of a community in a direct and ongoing way (\textit{n}=20, 43\%).\\

\textit{Type of Technology Developed}: Projects focused on a wide variety of technological artifacts and systems, including artifacts such as mobile or computer apps \cite{Dickinson2021-mv,Stowell2020-yf,Duarte2018-iq,Ssozi-Mugarura2016-on,Vigil-Hayes2021-rs,Khan2021-ug,Simm2014-rw,Molapo2016-nc,Thinyane2018-gt,Darroch2017-tm,Peters2018-js}, websites \cite{Keskinen2021-ce,Ginossar2010-uj,Tandukar2021-db,Grabill2003-ha,Sabiescu2014-ka}, robots \cite{Neto2021-jf,Moharana2019-ra}, and drones \cite{Wojciechowska2020-jv} as well as systems for knowledge management \cite{Winschiers-Theophilus2013-xj,Kapuire2015-ou,Blake2021-hg}, air quality monitoring \cite{Aoki2017-nl,Hsu2017-sh}, co-operative canoe paddling \cite{DiSalvo2012-rs}, among others.\\

\textit{Time Frame}: There was a similar number of projects transpiring over one or more years (long-term, \textit{n}=17, 36\%) and projects taking place for one month to one year (mid-term, \textit{n}=20, 43\%). Only nine projects (19\%) in the included articles reported on a period of less than one month (short-term). One article did not refer to the length of the project. The majority of `community-based research' projects were long-term (55\%); the majority of `participatory research' projects were mid- or short-term (78\%).\\

\section{Findings: Dimensions and Tensions of Community-Collaborative Projects}

This section continues our presentation of findings, focused on those items that emerged as part of our thematic analysis. We structure these findings according to seven themes of community-collaborative projects, which we have conceptualized as \textit{dimensions}. Dimensions are the elements or factors contributing to the holistic, participatory engagement of researchers and communities in a project.

Within each dimension, \textit{tensions} can emerge relating to the power and position of researchers and the computing research environment, relative to research participants and community members. We structure the findings to reflect the evolution of a project: from establishing the partnership at the project's outset, to enabling participatory models for decision-making and involvement in collaborative research activities, and finally, sustaining project results.

\subsection{Establishing a Partnership}

\subsubsection{Defining a Community Partner}

Identifying, selecting, and soliciting partners using CCA for research projects proved challenging for researchers without existing relationships with communities of interest. CCA encourage researchers to rely on self-definitions of `community' to identify partners \cite{Pine2020-zb,Winschiers-Theophilus2013-xj}. Expressions of self-identification may take a variety of forms, including (but not limited to) the existence of community-based organizations and places of assembly for members of the community (physical or virtual). Most researchers from the included articles partnered with an existing community-based organization or representative group to identify and connect with participants for their projects (\textit{n}=31, 66\%) (in some studies, the first group of participants acted as intermediaries, recruiting additional participants \cite{Winschiers-Theophilus2013-xj,Peters2018-js}). Some researchers were engaged directly by a community themselves \cite{Kapuire2015-ou,Keskinen2021-ce,Hsu2017-sh}, while others allowed individual participants to self-select by sharing advertisements through social media and posting in places of assembly for a community (\textit{e.g.}, invite-only chat groups \cite{Ahmed2021-ew}).

The status of participants was mixed across the reviewed projects. Most projects included `users', intended `users', or those who would maintain the artifact or system as research participants (\textit{n}=44, 94\%). Other members of a community---for example, caregivers for an individual participant or representatives of a community-based organization---were also often included as participants (\textit{n}=23, 49\%). Finally, a minority of projects included non-`users' or non-intended `users' of the artifact or system as research participants (\textit{n}=9, 19\%). For example, by co-designing a canoe to be used by paddlers with visual impairments in a public setting, Baldwin et al. \cite{Baldwin2019-eu} engaged `users' of the canoe (both visually impaired and sighted) and non-`users' from the general public in the design process.

A few projects also included representatives from the partner community on the research team. For example, Kapuire et al. \cite{Kapuire2015-ou} describe a seven-year project with a rural community in Namibia from the perspective of a researcher ``native'' to the community, to understand ``community gains'' from the project. Others included members of the partner community team as co-authors on research articles produced from the project \cite{Dickinson2021-mv,Grond2019-vw,McNaney2018-mv,Blake2021-hg}. 

\subsubsection{Building Relationships}

Papers in the corpus reported efforts to cultivate trusted relationships between researchers and partner communities prior to and during their projects. These relationships took time and effort to build on the part of both researchers and their partners. Most projects from the included articles involved some relationship-building activities with participants and/or community members prior to research activities, even when there was an existing relationship with the researchers (\textit{n}=34, 72\%). A minority of projects described researchers engaging with participants for the first time during research activities (\textit{n}=13, 28\%). Researchers without existing relationships with participants reported extensive efforts to build relationships to support a collaborative project. Such projects also often required efforts to build relationships between researchers and community members who were not participants in the project. Neto et al \cite{Neto2021-jf}, for example, developed relationships with many different members of a school before commencing a project with specific children and parents. Likewise Vigil-Hayes et al. \cite{Vigil-Hayes2021-rs} worked over 18 months to establish a community advisory board to guide researchers' engagements with research participants and the project overall. Many papers discussed ways researchers showed up in the setting of the community to build these relationships: as Ssozi-Mugarura et al. \cite{Ssozi-Mugarura2016-on} put it, through ``immersion into the culture of the community by the researcher to build trust and negotiate expectations.''

Coincidentally, engagement prior to commencing a research project is also reflected in recent research funding trends. For example, the US National Science Foundation's Smart and Connected Communities program ``encourages researchers to work with community stakeholders to identify and define challenges they are facing, enabling those challenges to motivate use-inspired research questions'' \cite{NSFSCC}. Likewise, community advisory boards are increasingly becoming a necessary and expected part of biomedical research in the US \cite{newman2011peer, NSFSCC}, Europe \cite{EURORDIS}, and various other countries \cite{manda2013establishing, zhao2019forming}.

Our analysis identified different approaches to initiating a project: some researchers reported relying on communities to identify problems to be addressed with technology (\textit{n}=14, 30\%); many researchers described using their own knowledge and experience to identify problems to be addressed with technology, doing so with some collaboration with a community (\textit{n}=33, 70\%). The latter was particularly common when community members had minimal experience engaging with computing or technology. When a project was initiated by a community, researchers and the partner community often had an existing relationship or researchers shared details about their knowledge of a problem with a community, who then sought a research partnership. For example, Duarte et al. \cite{Duarte2021-cz} presented at a summit at which a community shared interest in the problem space, and later invited the research team to the community site to plan a collaborative project. When a project was initiated by researchers, there was a mix between projects involving the community from the stage of planning research activities (\textit{n}=21, 45\%) and from the stage of conducting the activities (\textit{n}=12, 26\%).

Almost all projects were situated in the community (physically or in virtual community sites) for at least one stage of the project (\textit{n}=46, 98\%). Most were situated in the community throughout the project (\textit{n}=26, 55\%) or for multiple stages of the project (\textit{n}=18, 38\%). Harrington et al. \cite{Harrington2019-ze}, for example, intentionally situated design workshops within the senior village of their partner community: a low-income, predominantly African-American neighborhood of older adults. Similarly, Ahmed at al. \cite{Ahmed2021-ew} situated research with a virtual community through a chat server commonly used by the community.

\subsection{Enabling Participatory Models for Decision-Making and Collaborative Research}

\subsubsection{Integrating Community Knowledge and Practices}

Community knowledge and practices reflect the strengths and resources of a community, and include the skills or assets of individuals, networks of relationships, and mediating structures \textit{e.g.}, community centers \cite{Israel1998-yg}. The corpus of papers includes a mix of attempts to integrate community knowledge and practices with the design process. Many projects (\textit{n}=26, 55\%) reported on attempts to incorporate some aspects of community knowledge or practices with design processes for the project. Dickinson et al. \cite{Dickinson2021-mv}, for example, relied heavily on the network of street outreach workers managed by a partner organization to develop infrastructure for those workers to connect and collaborate. Some projects held research activities at common times and places of assembly for the community: \textit{e.g.}, regular meetings at a community center, alongside church services, or integrated with ceremonies \cite{DiSalvo2012-rs,OLeary2020-qw,Stowell2020-yf,A_Zewge2015-hr,Zhu2019-ak}. In a few cases, researchers reported on significant efforts to reorient the traditional paradigm of HCI research and deconstruct design processes to give primacy to community knowledge practice \cite{Winschiers-Theophilus2013-xj,Sabiescu2014-ka,Le_Dantec2015-zk}. In Sabiescu et al \cite{Sabiescu2014-ka}, for example, local customs and values determined the format and flow of the design process, including who could be involved, how decisions were made, and how consensus was achieved.

Most of the included articles described collaborative ideation activities between researchers and participants (\textit{n}=44, 94\%). Notably, however, there were few examples of researchers and communities collaborating to conceptualize a \textit{problem} to be addressed through the development of a technological artifact or system. Community partners were mostly included in ideating on aspects of a problem conceptualized by researchers prior to the project or ideating solutions to a problem entirely conceptualized by researchers.

Many of the articles also included participants in developing prototypes (\textit{n}=26, 55\%) and critiquing prototypes or final designs (\textit{n}=23, 49\%). Involving participants in prototyping and critiquing may serve to demonstrate that researchers align with the partner community’s goals for the final product or service \cite{Baldwin2019-eu}, but has limitations: ``usability'' and other critiques of solutions to a problem conceived by computing researchers does not itself reflect relevance or value of the solution to a community \cite{Unertl2016-gr}. Involving participants in creative or technical tasks requires researchers to focus on providing support for non-expert participants, and requires a good deal of effort from community members \cite{McNaney2018-mv}, increasing the importance of ensuring the solution meets the needs of the community.

\subsubsection{Sharing Control}

A key characteristic of the corpus was the shared control of multiple aspects of the project among researchers and communities. Sharing control requires researchers to think of the community as the authority on their community---their ``domain expertise'' \cite{Harrington2019-yc}. In most articles, researchers and communities shared control of planning and managing data collection and research activities (\textit{e.g.}, questions for interviews, format and context for workshops) (\textit{n}=33, 70\%). Other articles also shared control and decision-making during the development of the final artifacts or systems (\textit{n}=27, 57\%), with a strong overlap with articles including participants in prototyping activities. Notably, few papers discussed sharing control for data analysis (\textit{n}=4, 8\%). Finally, two articles discussed the concept of ``scaffolding'' leadership during the project---with researchers supporting communities to lead some aspects of the research process---as an alternative to sharing control of individual research activities \cite{Peters2018-js,Blake2021-hg}.

The majority of included articles described efforts by researchers to attend to power differentials between researchers and communities by elevating the voice of communities (\textit{n}=27, 57\%). Some articles emphasized the importance of researchers providing assurances to participants that their community knowledge is of equal value to the collaboration as the knowledge of researchers \cite{Ssozi-Mugarura2016-on,Keskinen2021-ce,McNaney2018-mv,Moharana2019-ra,Grabill2003-ha,Fox2014-ko,Stillman2012-dd,Zhu2019-ak}. In projects involving people with a disability, researchers noted the considerable effort required to adapt methods to enable and support participation \cite{McNaney2018-mv,Grond2019-vw,Neto2021-jf}. Neto et al \cite{Neto2021-jf}, for example, created a multi-sensory toolkit to enable the participation of visually impaired children in prototyping tasks simultaneously with their classroom peers.

\subsubsection{Cycles of (Re)Production and Adaptive Collaboration} The projects described in the paper corpus employed varied research techniques, including ethnography, surveys, interviews, focus grou-ps, and design workshops. Hsu et al. \cite{Hsu2017-sh} encourage such flexibility in approach, and warn against the temptation to work with a community in a single implementation phase to replicate a successful project from another community setting. Consistent across most projects, however, was a multi-phase process with community partners to iteratively develop artifacts and systems. Almost all projects reported multiple phases of involvement with a partner community, focused on a type of activity or goal (\textit{n}=31, 87\%), instead of a single phase.  The paper corpus suggests, then, that allowing space for researchers and communities as partners to connect, reflect, and re-connect reveals, to researchers, the challenges unique to a community and, to a community, the constraints of technology design.

The finding above is not surprising given the centrality of iteration to many HCI research methods \cite{Dourish2020-zc}. Less evident in the paper corpus, however, was reporting on the demands placed on communities by iteration. While cycles of production and reproduction may enable the transfer of knowledge between researchers and community partners, such cycles also increase the commitment required of community members to participate in collaborative projects relative to other research approaches. As Dourish et al. note \cite{Dourish2020-zc}, many marginalized communities experience short-term, periodic and iterative engagements with outsiders in various aspects of their lives---\textit{e.g.}, public policy development---with longer periods of neglect outside those engagements. HCI research must consider the broader cycles in which projects are embedded, and the ``affective demands'' of those cycles on community partners \cite{Dourish2020-zc}.

Beyond multiple phases of involvement, many projects also referred to the need to adapt the research methodology at some stage of a project (\textit{n}=31, 66\%). The methodology invariably shifted as the partners began to collaborate, including those projects starting with a preconceived methodology. In fact, some researchers planned explicit stages in a project for community partners to change or reconfigure the plan and goals of activities, or the project overall \cite{Ahmed2021-ew,Fox2014-ko,Harrington2019-ze}. Other examples of adaptive research planning included incorporating a high degree of flexibility in the format of design workshops, to encourage participation and ``participant-led'' design \cite{Wojciechowska2020-jv}; creating space in the design process for open activities such as storytelling \cite{Dickinson2021-mv}; and fitting research activities around the everyday ``dynamics and rhythm'' of local residents’ lives \cite{Winschiers-Theophilus2013-xj}. However, some authors noted that computing researchers must translate community contributions into research contributions such as academic papers and technology designs, which computing researchers need to produce to secure ongoing resources to continue the work. This necessary production of particular artifacts can limit the openness of researchers to allow a community's experience to guide methodology \cite{Le_Dantec2015-zk, Duarte2021-cz}

\subsection{Sustaining Results of a Project}

\subsubsection{Ensuring Mutual Benefits}

Most articles identified outputs of the research project for both researchers and the community (\textit{n}=29, 62\%) or outputs for the community, with outputs for researchers limited to publication (\textit{n}=11, 23\%). A minority of projects identified only outputs of the project for researchers (\textit{n}=7, 15\%). These projects often involved the development of design principles for a technological artifact or system immediately relevant for researchers, but not for the community \cite{OLeary2020-qw,Wojciechowska2020-jv,Haimson2020-ip}.

The majority of papers also reported on attempts to balance benefits from the projects between both researchers and the partner community (though not necessarily similar benefits across them). Around half of the included articles identified benefits of a project to individual participants and to the community as a whole (\textit{n}=25, 53\%). Fox and Le Dantec \cite{Fox2014-ko} further encourage researchers to make explicit the benefit of each step in the research process to individual participants. Others identified only benefits to the community as a whole (\textit{n}=13, 28\%) or individual participants (\textit{n}=7, 15\%). Two articles did not specify benefits for individuals or the community as a whole.

Benefits for each project differed for individual participants and, in particular, for communities. One benefit for both participants and communities identified across many projects was `capacity building' (\textit{n}=31, 67\%), particularly with respect to improvements in the capacity of the community to advocate for themselves. For example, some articles reported participants using technical skills developed during a project (\textit{e.g.}, data collection and data management) to inform efforts to advocate on issues related to technology or public policy \cite{Aoki2017-nl,Ssozi-Mugarura2016-on,Shilton2008-jk,Xu2017-cg,Blake2021-hg,Hsu2017-sh}. A community-based organization may also receive recognition from community members for addressing a problem relevant to the community. Given this potential benefit, Le Dantec and Fox \cite{Le_Dantec2015-zk} encourage researchers to situate contributions in published work in terms of the partner community instead of the research community.

Most articles described one benefit of CCA as developing researchers’ understanding of a partner community and their context (\textit{n}=39, 83\%). Researchers argued such projects gave them a better understanding of `users' (\textit{e.g.}, their experiences, challenges, needs) and the integration of technology into the communities in which it will exist \cite{Tandukar2021-db,Shilton2008-jk,Duarte2018-iq}. Reflecting on the similarity of challenges faced by their partner community and other marginalized communities, Haimson et al. \cite{Haimson2020-ip} suggest that the understanding of a community gained by researchers from a community-collaborative project may be relevant to understanding the experience of other marginalized communities.

Our review found that outputs and benefits of projects for participants and communities are often incremental rather than immediate. While researchers may receive immediate information and feedback on design requirements, for communities the outputs of the project (\textit{e.g.}, technology designed for their unique circumstances) are received much later, if at all. Equally, while building skills and capacity for advocacy may be highly valued by individual participants and communities, they are benefits typically experienced in future contexts and may be unevenly experienced throughout the community. On the other hand, a deeper understanding of the communities in which technology will exist may be immediately useful to researchers to tailor deployment plans.

In addition, few papers in the corpus described approaches to measuring the benefits of projects for communities, either immediately following a project or in an ongoing way. The lack of measurement of benefits for communities may also indicate a challenge in bridging the measurement approaches researchers know and regularly use with the benefits community partners may derive from a project. As noted in Duarte et al. \cite{Duarte2021-cz}, researchers are in general measured by productivity, whereas project partners may be `measured' by the stabilization of relationships in a community.

\subsubsection{Sharing Results and Maintaining Relationships}

The included articles encourage researchers to communicate the results of a project alongside the community. This principle is reflected in the practice of CBPR in public health, which encourages community input into plans for dissemination, dissemination of results to the community, and results translated to social action \cite{Chen2010-ib,Unertl2016-gr}.

Around one-third of projects involved researchers partnering with participants to share the results of the research, with other members of the community or to a wider set of stakeholders (\textit{n}=18, 38\%). In DiSalvo et al. \cite{DiSalvo2012-rs}, for example, researchers organized a public event modeled on a science fair, enabling participants to present designs to their community and other stakeholders. Others noted steps the researchers took to share results with participants (\textit{n}=18, 38\%), though some papers did not report such steps (\textit{n}=11, 23\%). Whether these steps were absent in practice or not, however, cannot be confirmed based on publications alone.

A majority of papers also considered the sustainability of the artifact or system in some way (\textit{n}=31, 66\%). In many cases, researchers considered how the community would make ongoing use of the artifact or system. Vigil-Hayes et al \cite{Vigil-Hayes2021-rs}, for example, included an explicit research question to identify design elements and community factors that would help the partner community engage with an app in a sustainable manner. Ahmed et al \cite{Ahmed2021-ew} argue researchers and practitioners must create ``sustainable structures'' during the project for the community to own and control the work, though others noted that researchers need to consider how the structures will be funded ongoing \cite{Ginossar2010-uj,Simm2014-rw}. Others provided training and education to the community during the project on operating and managing an artifact or system \cite{Baldwin2019-eu,Hsu2017-sh,Xu2017-cg,Duarte2018-iq,Ssozi-Mugarura2016-on}.

Few articles, however, reported on what and how relationships with a partner community would be maintained after researchers concluded their substantive engagement. As Ginossar \& Nelson \cite{Ginossar2010-uj} note, community-based organizations may change or dissolve, requiring ongoing work by researchers to monitor and maintain community relationships.

\section{Discussion}

In this section, we first reflect on the findings of our review and the structural changes necessary in HCI---and computing research more broadly---to achieve the aims of CCA. Alongside these structural changes, we offer four methodological proposals to move our field closer to centering communities in research. Finally, we encourage ongoing reflexive practice by researchers conducting projects based on CCA, to develop a `community of practice' in our field around CCA.

\subsection{Structural Changes}

A key feature of CCA, such as AR and CBPR, is the involvement of community stakeholders in \textit{all} stages of a research project. Our review found, however, that most computing research publications drawing on CCA report on the involvement of community stakeholders in more piecemeal ways. First, the review found most studies involved community stakeholders after a problem was defined. Researchers tended to rely on their own knowledge to conceptualize a project to be later delivered in partnership with a community. Second, there were few examples of researchers sharing control of data analysis with communities. Some authors reported examples of researchers sharing control with community members for collecting and curating data \cite{Xu2017-cg,Halfaker2020-jp, Hsu2017-sh}. Sharing control over data collection and curation presents an opportunity for social and political action by participants, through which ``big secondary data'' may be integrated and verified with community perspectives \cite{akom2016youth, Hsu2017-sh}, though, at the same time, Pierre et al. \cite{Pierre2021-rs} warn that researchers must consider the ``epistemic burden'' to marginalized communities of this data work. In most studies reviewed for this paper, however, researchers were primarily or exclusively responsible for data analysis. This finding reflects Braten's concerns around ``model monopoly'' \cite{braten1973model}, which describes the tendency for professional researchers to dominate discussion, particularly during more traditionally scientific activities like data analysis.

AR practitioners argue these two stages---conceptualizing a research problem prior to solutions development and analyzing data---are the most critical to involve the community to achieve the aims of AR \cite{stringer2007action}. Community involvement in identifying issues and defining questions for computing research is critical for ensuring projects address issues relevant to communities. Equally, involving community members in the analysis of data produced through research activities---consuming, digesting, and engaging with data rather than only producing or being the subject of data---builds the capacity of communities to address their own issues.

The limited involvement of communities in these research stages across the included articles suggests a systemic disconnect between the theory and the practice of CCA as reported in computing research. The findings reveal the choices HCI researchers invariably make in planning projects based on CCA: prioritizing tightly-scoped projects, community involvement only during delivery of research activities and researcher-led data analysis. While such choices may reflect researcher interests to attract funding and resources or publish results, rather than community interests, they are choices made within the social systems and organizational structures that make up the current paradigm of HCI research. We argue, therefore, that structural reform to the research and publication of CCA is required to address the systemic disconnect. We ask: how could funding in HCI, and computing more broadly, support engagement with communities to conceptualize research projects, rather than specifying an outcome for researchers to achieve from a project? How could publication processes encourage reporting focused on the development of community members’ skills in data analysis and technology development, rather than the validity and reliability of experimental results? From this study, we hope to catalyze a research agenda to study the systemic issues that restrict HCI research, and computing more broadly, from achieving the goals of AR, CBPR, and other CCA.

Additionally, before developing any technological artifact or system, CCA encourage researchers and communities, as a group, to collaboratively develop vision and operational statements \cite{stringer2007action}. \textit{``Vision statements allow the entire team to work together to decide what the issues are and how all of the concerns of the people involved will be accounted throughout the process''} \cite{Hayes2011-be}. Notably, these vision statements rarely emerge directly from a community–researcher initial discussion. They tend, instead, to emerge from discussions around empirical data in the form of field notes, survey responses, transcripts of focus groups and interviews, and so on. These empirical data may be collected by community members, researchers, or both. The space and norms of an academic publication may or may not leave space to describe how these came about. Rather, one might see an empirical paper based on ``formative research'' published with limited mention of participants followed by a design paper that appears to have a problem statement pre-emptive to the collaborative project. New forms of reporting on such research projects would have to be normalized and supported to be able to understand the entirety of the engagement.

\subsection{Methodological Proposals}

In the sections below, we suggest four areas for methodological innovation in HCI research, which we propose to be investigated alongside the broader research agenda referred to in the section above.

\subsubsection{Defining Communities}

A conception of `community' as an aspect of \textit{both} collective and individual identity is at the heart of CCA \cite{Xu2017-cg}. This definition suggests that social relations between community members are key to the existence of a community, in addition to any common identifying characteristics of individuals. Selecting a group of individuals to engage in a project based on characteristics they have in common---from the perspective of a researcher---would not align with the conception of a community in projects drawing on CCA. This approach also ignores the important observation that communities are not only defined by commonalities among members, but just as readily by their differences, which create vital spaces for open disagreement and dissensus about areas of mutual \textit{value} to community members \cite{DiSalvo2011-ca}. To our knowledge, communities are often defined in HCI and computing research by grouping the behaviors and attitudes of individuals' interactions, often with products and services. This approach (focusing only on individual identity and not on collective identity) can disempower the group of individuals relative to the researchers during a project based on CCA. Defining a community is an act of power, and for marginalized communities---where conditions are often set by people or groups outside the community---defining their community on their terms may be an act of reasserting power \cite{Hovmand2014-oj}.

Most authors of the included articles addressed the challenge of identifying a partner community by engaging with existing community-based organizations as partners in research or engaging with existing places of assembly for members of the community. However, as community membership and gathering moves online, we will see increasingly decentralized, distributed ways of connecting, such that locality for community could become more difficult to define and stabilize. An opportunity for methodological research in HCI, then, is to develop processes for identifying communities in online networks of `users' and/or other stakeholders, facilitating connection between community members, and supporting the representative infrastructure necessary for community members to engage---as a unit---in research.

\subsubsection{Facilitating Contestability of HCI Research and Design Practice}

Rorty argued that science can be used to open up new conversations and keep them going, even through conflict, aiming for debates and arguments to bring people into ``communicative clarity” \cite{rorty1979philosophy}. This kind of approach may become muddled when researchers focus on building consensus or agreement on how to move a project forward to distill outcomes for scientific publication. Such consensus-based, outcome-focused processes can obscure the difficult debates and arguments that arise in the process of consensus building, at times obscuring or eliminating them altogether. Indeed, discussion, debate, and communication among all stakeholders are fundamental elements to the generation of knowledge and the production of change in CCA, but academic publishing norms currently leave little space for such elaboration. 

Here, HCI has a unique opportunity as an interdisciplinary field \cite{Blackwell2015-yg} categorized by dissensus \cite{Wright2015-uu}. These features of HCI have enabled the field to challenge assumptions in computing (\textit{e.g.}, around 'users' and usability), which we can now build upon to further support notions of engagement and participation by and between researchers and community members, all of whom may variably agree or disagree on a wide variety of concerns for a project. The articles included in this review focused on consensus-building between researchers and communities through collaboration and iteration in the delivery of a project (\textit{e.g.}, to develop tools to use for advocacy, or as a tool for the community). In some ways, this finding mirrors recent critiques that HCI can be prone to ``interest convergence'', in which concessions to inclusion require benefits to those in power \cite{Ogbonnaya-Ogburu2020-sk}. One is left to ask: how can HCI researchers engage with communities prior to research projects (long-term, ongoing collaboration) to build the confidence and capability of a community to critique HCI research and design practice, or even specific technologies \cite{Sabiescu2014-ka}? Perhaps just as importantly, how can researchers who have conceived of such participatory and collaborative projects with community partners describe that success in ways that academic venues can comprehend and respect? When HCI researchers act as facilitators between communities and technology researchers/designers, creating forums for partner communities to \textit{contest} existing technology designs and embodiments with researchers, and encourage alternative approaches, they are creating the conditions for dissensus and debate that Rorty, as a pragmatist, argued for \cite{rorty1979philosophy}. 

\subsubsection{Building from Community Knowledge and Practices}

Our review found that researchers often pivoted during projects to allow community experience to guide methodology, though reporting on the challenges of doing so was limited. Additionally, while our review found many examples of researchers incorporating some aspects of community knowledge and practices in the delivery of projects, only a few articles reported significantly reorienting research and design practices based on community knowledge and practices. More commonly, researchers adapted methods to a community setting (see \textit{e.g.}, \cite{Le_Dantec2015-zk,Harrington2019-ze}), but the methods remained firmly embedded in HCI and/or computing research and practice. As Winschiers-Theophilus \& Bidwell \cite{Winschiers-Theophilus2013-xj} suggest, the practice of translating local knowledge and practices into research tools and processes can serve to challenge and refine existing HCI concepts and theories.

Two opportunities appear given these findings: for the HCI community to make space for reporting on the process and challenges of pivoting to allow community experience to guide research methodologies; and for HCI researchers to draw on their engagements with communities in CCA projects to reflect on and improve HCI research and design methods. Methodological innovation could guide researchers to an emphasis on documenting and sharing community knowledge and practices that may be useful for computing research, and translating user experience and design knowledge and practices to broader communities with whom we engage.

\subsubsection{Sustaining Results}

Since at least the 1970s, researchers have been discussing whether and how community members benefit from CCA \cite{green1973unresolved}. Our review demonstrates that this issue has not diminished in the decades since, as we found limited guidance in computing research on how to define and measure the benefits of projects for communities as a whole as well as for individual participants. We will likely see increased risk for unbalanced relationships and benefits between researchers and communities as the complexity of technologies and the challenges of society they seek to address increase, in tandem.

This finding points to the need for methodological research in HCI to bridge the gap between metrics relevant to researchers and those relevant to communities. New or different project evaluation metrics may require a shift to: 1) qualitative measures of success for researchers and the community (\textit{e.g.}, measures of subjective well-being and identity); 2) measures emphasizing stability in relationships between researchers and the community; 3) measures emphasizing stability in relationships between a community organization and its members, or among community members; and 4) the research profile or profitability of the researchers' organization relative to the benefits experienced by the community.

Furthermore, few of the included articles reported on evaluations conducted over time to assess the sustainability of a project, which is inextricably linked to a community realizing the benefit of a project. This points to the need to develop processes and incentive structures within the scholarly HCI community for ongoing evaluation and reporting on project outcomes (generally, and specifically for communities) past the point of novelty. For example, researchers could build time into the project plan to ensure both funding and expectations make room for reaching back out to the community and conducting ongoing assessments. The scholarly HCI community could also consider incentives such as specific publication tracks for ongoing reporting of project and community outcomes, and the creation of new spaces to bring HCI researchers doing work with broader communities together to learn from one another.

\subsection{Reflexive Practice}

Interrogating researcher stance is fundamental to CCA as outlined in discussions around CBPR \cite{Simmons2019-lu}, AR \cite{Hayes2011-be}, and community-institution partnerships \cite{gregoire2007ethics}. Reflecting on their own engagements, Le Dantec and Fox argue a key reason for adopting CCA from other disciplines into computing research is to ``embrace'' subjectivity as researchers and account for this subject position in the social context of technology development \cite{Le_Dantec2015-zk}. CCA methods encourage the notion of a `friendly outsider' role for researchers and explicitly rejects the idea that they should distance themselves from community partners, thereby rejecting the traditional roles of `subject' or `participant' for a strong role as a collaborator, leader, or partner in the research.

The exact role to be created for community members is variable depending on the context of the project, but the role for the researcher in these approaches is clearly intended as a reflective, engaged participant and not a distanced scientist. Liang et al. \cite{Liang2021-qe} identify tensions for HCI researchers working with marginalized groups that may guide such reflections, focusing on the nature of engagement with community partners (whether engagement is extractive or tokenizes identities), membership of researchers in the community, disclosure of researcher stance, and the nature of allyship in the pursuit of social change.

Considering researcher stance should also include awareness of the historical relationship between a community and research institutions \cite{Pine2020-zb}. Prior to commencing a community-collaborative project, researchers may need to consider the trust (or lack thereof) held by the community in their organizations. Addressing a lack of trust is an issue affecting many researchers and research organizations more broadly than in the conduct of CCA, nonetheless it remains a barrier to consider in the delivery of any community-collaborative project.

Previous mistreatment of people in research and technological innovation \cite{scharff2010more, Thomas1991-yg, Pacheco2013-te, Cochran2008-fx}, and experiences of racism and other forms of discrimination by dominant social and cultural institutions, have affected trust in research and technology development more broadly, and present a complex historical backdrop that could impact whether and how a community decides to participate in a computing research project. For projects involving the collaborative development of algorithms, for example, a lack of trust in the existing social or economic system of intervention contributes significantly to low comfort in algorithmic decision-making \cite{Brown2019-to}. Researchers conducting such projects should consider the trust (or lack thereof) held by the community in the social or economic system a project relates to, in addition to the trust held by the community in the researchers' organization, as researchers may be perceived by the community as a representative of the broader system.

Interrogating researcher stance should involve reflecting on the socio-cultural and political environment for the community, and documenting tensions and conflicts occurring during a project \cite{Erete2018-qr}. It also includes documenting the work required to build and maintain relationships throughout the project, to ``make visible the work before the work, and the work to keep the work going” \cite{Le_Dantec2015-zk}. One effective method to achieve this documentation may be auto-ethnographic field notes \cite{Ahmed2021-ew}, though protocols around sharing these notes within the project team and researchers more broadly would need to be defined for each project.

Furthermore, an important goal of documentation, and protocols around it, is to increase transparency in the research process more broadly. Not only do the individual decisions and outputs of a participatory process matter, but also the ``who”, ``how” and ``why” behind them \cite{Bell2012-lm}, all of which are important antecedents to critical, systematic reflection in participatory projects \cite{Frauenberger2015-bj}. Overall, the process of documenting experiences and reflections could assist the development of a community of practice in HCI around CCA.

\section{Conclusion}
We conducted a systematic literature review and reflexive thematic analysis of CCA to computing research. Drawing on the literature, we note there are seven dimensions of community-collaborative projects contributing to holistic, participatory engagement of researchers and community partners, and tensions can emerge related to the position of computing researchers, and the computing research environment, relative to community partners. Our intent is not to resolve the tensions through this review. There are and will remain different incentives and significant power imbalances among computing researchers, interdisciplinary teams, and communities. The structural conditions of computing research, including HCI---institutional cultures and publication conventions---also limit the ability of researchers to conduct and report on projects in ways that reflect the aims and benefits of CCA.

While the articles reviewed for this paper demonstrate motivation and progress among computing researchers doing this type of work, there is a need to study and address the systemic issues impacting our ability to fully achieve the goals of CCA. We hope that, along with certain methodological innovations, this research agenda will enable responsible and mutually-beneficial projects based on CCA to become a central component of HCI (and computing research more broadly) in the coming years. 







\begin{acks}
Our sincere thanks to Alex Zafiroglu, Carl DiSalvo, and anonymous reviewers for valuable feedback on the paper.
\end{acks}

\bibliographystyle{ACM-Reference-Format}
\bibliography{main.bib}










\end{document}